\documentclass[pra,superscriptaddress,,nofootinbib]{revtex4}
\usepackage{graphicx,amsmath,amsfonts,algorithm}
\usepackage{graphics}
\usepackage{amssymb}
\usepackage{enumitem}
\usepackage{color}
\usepackage{bbm}
\usepackage{subfigure}

\def\ket#1{| #1 \rangle}

\newtheorem{prop}{Proposition}\def\PRO{\begin{prop}}\def\ORP{\end{prop}}
\newtheorem{coro}{Corollary}\def\COR{\begin{coro}}\def\ROC{\end{coro}}
\newtheorem{theo}{Theorem}\def\TH{\begin{theo}}\def\HT{\end{theo}}
\def\TH{\begin{theo}}\def\HT{\end{theo}}
\newtheorem{defi}[prop]{Definition}\def\DE{\begin{defi}}\def\ED{\end{defi}}
\newtheorem{lemme}[prop]{Lemma}\def\LE{\begin{lemme}}\def\EL{\end{lemme}}

\def\EQ#1{\begin{eqnarray}#1\end{eqnarray}}

\def\op#1{\hat{#1}}

\def\op#1#2{|#1\rangle\langle#2|}

\def\dm#1{\op{#1}{#1}}

\newcommand{\djj}{d\kern-0.4em\char"16\kern-0.1em}

\begin{document}

\title{Realization of Quantum Digital Signatures without the requirement of quantum memory}

\author{Robert J. Collins}
\affiliation{SUPA, Institute of Photonics and Quantum Sciences, School of Engineering and Physical Sciences, David Brewster Building, Heriot-Watt University, Edinburgh EH14 4AS, UK}

\author{Ross J. Donaldson}
\affiliation{SUPA, Institute of Photonics and Quantum Sciences, School of Engineering and Physical Sciences, David Brewster Building,  Heriot-Watt University, Edinburgh EH14 4AS, UK}

\author{Vedran Dunjko}
\affiliation{SUPA, Institute of Photonics and Quantum Sciences, School of Engineering and Physical Sciences, David Brewster Building,  Heriot-Watt University, Edinburgh EH14 4AS, UK}
\affiliation{School of Informatics, Informatics Forum, University of Edinburgh, 10 Crichton Street, Edinburgh EH8 9AB, UK.}
\affiliation{Laboratory of Evolutionary Genetics, Division of Molecular Biology, Ru\djj er Bo\v{s}kovi\'{c} Institute, Bijeni\v{c}ka cesta 54, 10000 Zagreb, Croatia}
\affiliation{Now at: Institute for Quantum Optics and Quantum Information,
Austrian Academy of Sciences, Technikerstr. 21A, A-6020 Innsbruck, Austria}

\author{Petros Wallden}
\affiliation{SUPA, Institute of Photonics and Quantum Sciences, School of Engineering and Physical Sciences, David Brewster Building,  Heriot-Watt University, Edinburgh EH14 4AS, UK}

\author{Patrick J. Clarke}
\affiliation{SUPA, Institute of Photonics and Quantum Sciences, School of Engineering and Physical Sciences, David Brewster Building,  Heriot-Watt University, Edinburgh EH14 4AS, UK}
\affiliation{Now at: School of Instrumentation Science and Opto-electronics Engineering, Beijing University of Aeronautics and Astronautics, 37 Xueyuan Road, Haidian District, Beijing 100191, China}

\author{Erika Andersson}
\affiliation{SUPA, Institute of Photonics and Quantum Sciences, School of Engineering and Physical Sciences, David Brewster Building, Heriot-Watt University, Edinburgh EH14 4AS, UK}

\author{John Jeffers}
\affiliation{SUPA, Department of Physics, John Anderson Building, University of Strathclyde, 107 Rottenrow, Glasgow G4 0NG, UK}

\author{Gerald S. Buller}
\affiliation{SUPA, Institute of Photonics and Quantum Sciences, School of Engineering and Physical Sciences, David Brewster Building, Heriot-Watt University, Edinburgh EH14 4AS, UK}

\begin{abstract}
Digital signatures are widely used to provide security for electronic communications, for example in financial transactions and electronic mail. Currently used classical digital signature schemes, however, only offer security relying on unproven computational  assumptions. In contrast, quantum digital signatures
offer information-theoretic security based on laws of quantum mechanics~\cite{QDS, ErikaOrig, OurNatComm}. Here, security against forging relies on the impossibility of perfectly distinguishing between non-orthogonal quantum states. A serious drawback of previous quantum digital signature
schemes is that they require long-term quantum memory, making them unfeasible in practice. We present the first realization of a scheme~\cite{OurArXiv} that does not need quantum memory and which also uses only standard linear optical components and photodetectors. In our realization, the recipients measure the distributed quantum signature states using a new type of quantum measurement, quantum state elimination~\cite{stevebook, OppenUSE}. This significantly advances quantum digital signatures
as a quantum technology with potential for real applications.
\end{abstract}

\maketitle
Digital signatures are used to ensure that messages cannot be forged or tampered with. Signed messages are also transferable, meaning that  it is unlikely that one recipient accepts a message as genuine, while another recipient, to whom the message is forwarded, rejects it. This important property is also called non-repudiation; a sender cannot deny having sent a message. Digital signature schemes are different from encryption, which guarantees the privacy of a message. Both are important cryptographic tasks. Quantum key distribution (QKD)~\cite{QKDreview, QKDReviewSecurity} can be used to distribute a secret key for information-theoretically secure encryption, and commercial systems are already available~\cite{RealQKD, TokyoQKD}. Analogously, digital signature schemes relying on quantum mechanics~\cite{QDS, ErikaOrig, OurNatComm, OurArXiv} can also
be made information-theoretically secure, in contrast to currently used classical digital signature schemes.
In this work we show that quantum digital signature (QDS) and QKD require similar experimental components and a comparable level of experimental complexity.

Protocols for quantum digital signatures have a
distribution stage and a messaging stage. We will describe the case with one sender and two recipients, but this can be extended to more recipients.
In the distribution stage, the sender, Alice, transmits quantum signature states to the recipients, Bob and Charlie. She chooses a sequence of $L$ states for each possible message that she might later want to send, for a suitable chosen integer $L$, and distributes one copy of each state sequence to each recipient. The quantum states are randomly chosen from a set of non-orthogonal states, in our realization four coherent states
$|\alpha\rangle, |\alpha e^{i\pi/2}\rangle, |\alpha e^{i\pi}\rangle$ and $|\alpha e^{3i\pi/2}\rangle$, with known magnitude $\alpha$.
The chosen phase sequences are analogous to a private key, known only to Alice. In the simplest case, to send a one-bit message later on, Alice distributes two sequences of states to each of Bob and Charlie, one corresponding to the possible message ``0", and one corresponding to the message ``1".

In the subsequent messaging stage, Alice accompanies the message she sends with the classical information about the corresponding sequence of quantum states; in our realization, the sequence of phases. A recipient of a signed message tests that this agrees with the previously distributed quantum signature states, and accepts the message as genuine if there are sufficiently few mismatches for the whole sequence. Similarly, to forward a message, a recipient forwards the message together with the  information about the corresponding quantum signature states. The new recipient again tests for mismatches and verifies if these are few enough.

Previous QDS schemes~\cite{QDS, ErikaOrig, OurNatComm} required that recipients store the signature states in long-term quantum memory until the messaging stage. Once a recipient is given the private information about a signature state, say, that it should be equal to some state $|\phi\rangle$, the best way to test for a mismatch is to make a quantum measurement with measurement operators $|\phi\rangle\langle\phi|, \mathbf 1 -|\phi\rangle\langle\phi|$. That is, to test if the state has any component orthogonal to the state it is declared to be.

\begin{figure*}
\label{Fig:setup}
\includegraphics[
height= 8 cm]{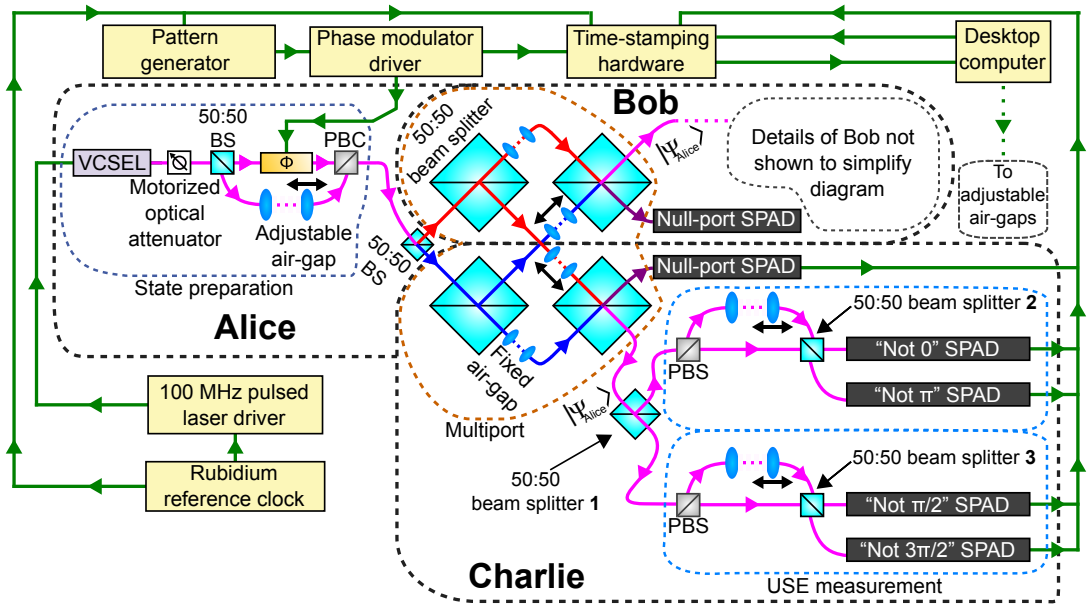}
\caption{Experimental setup for distributing
quantum digital signatures.
VCSEL denotes a vertical cavity surface emitting laser.
Alice uses a LiNbO$_3$ phase modulator to apply a phase shift $\Phi$, randomly chosen as 0, $\pi/2$, $\pi$ or $3\pi/2$, to each coherent state.
The recipients Bob and Charlie use an all-optical fiber multiport to ensure non-repudiation and guard against forging, consisting of the four beam splitters within brown dashes. For detection, the setups
within light blue dashes are used to eliminate one or more
possible phases. The detectors are silicon single-photon avalanche diodes (SPADs).
PBC denotes a polarization beam combiner and PBS denotes polarization beam splitters.
\label{Fig:setup}}
\end{figure*}

The requirement for quantum memory is clearly unfeasible at present. There may be days, weeks, or longer between the distribution and the messaging stages, whereas state-of-the-art quantum memories cannot achieve coherence times longer than tens of minutes at room temperature~\cite{qmem, qmemThewalt}. A protocol that circumvents quantum memory was suggested in~\cite{OurArXiv}, and our current experiment realizes a variant of this scheme. Here, the recipients measure the signature states directly at the end of the distribution stage. Only classical information needs to be stored. In~\cite{OurArXiv}, unambiguous state discrimination measurements were envisaged. In our realization, we improve on this idea so that Bob and Charlie instead use {\it unambiguous quantum state elimination}~\cite{stevebook, OppenUSE} to probabilistically exclude one or more phases for each signature state.

Our experimental setup is shown in Fig. \ref{Fig:setup}.
The state elimination measurement can for coherent states be realized using linear optics and photodetectors.
Each recipient uses two detection systems,
shown within dashed light blue lines in Figure \ref{Fig:setup}, where the signature states are interfered with reference pulses of phase 0 in the top and phase $\pi/2$ in the bottom interferometer.
Polarization routing~\cite{Marand} is employed for the orthogonally polarized signal and reference pulses, using the polarizing beam splitter
(PBS) and combiners (PBC).
The reference pulses enter through the left-hand input ports of
beam splitters ``2" and ``3" while the delayed signal pulses enter through the top.
Detecting photons in any of the
output ports
excludes one possible phase, similar to a recent realization of unambiguous state discrimination (USD)~\cite{ExpUSD}. Whereas USD
requires excluding all but one of the quantum states, we only require elimination of at least one state (phase).  This significantly increases the number of usable signature elements by requiring fewer detection coincidences.  The process of USE is summarized here and described in more detail in the Appendix \ref{Supp}.
To estimate
the resulting advantage, assume that the
amplitude entering Bob's and Charlie's measurement setups is $|\beta|^2$, and neglect e.g. phase imperfections.
The probability of excluding the coherent state of opposite phase to the one that is sent is then $1-\exp(-|\beta|^2)=p$, and the probabilities to exclude the other two are $1-\exp(-|\beta|^2/2)=q$.  The probability of excluding all three states that were not sent is $pq^2$, while the probability of excluding at least one of them is $1-(1-p)(1-q)^2$,
which is always greater.  If, as in our experiment, $|\beta|$ is small, then this quantity is much greater than $pq^2$.

A forger must avoid declaring a phase that has been eliminated; more precisely, avoid this for sufficiently many signature sequence positions. If Bob (or Charlie) succeeds in eliminating three of the four possible phases for one signature position, then a forger must select the single remaining phase to avoid a mismatch. But even if just one phase is ruled out, a forger must avoid selecting this phase.
With USE, therefore, many events which would count as non-detected if using USD will now contribute to the detection of forging. Consequently, using state elimination
leads to an improvement in the signature generation rate.
For both USE and USD, the forger's probability of avoiding too many mismatches
decays exponentially with the signature length  $L$.

A more detailed security analysis is found in \cite{OurArXiv} and in the Appendix \ref{Supp}.
We have examined security for a single use of the protocol, for general repudiation attacks and all forging attacks except those involving entangling operations on successive signature sequence states. So-called ``composable security" remains an important issue.
In short, security against forging follows since Alice's signature states are chosen from a set of non-orthogonal quantum states, which cannot be distinguished perfectly. Only Alice has the full
description of these states.
Note that the number of recipients depends on protocol parameters, since if too many copies of Alice's signature states are available, or if $|\alpha|^2$ is too large, then
the private phases could be determined reliably enough to forge a message unless protocol parameters such as $L$ are adjusted.

\begin{figure*}
\includegraphics[
height= 7 cm]{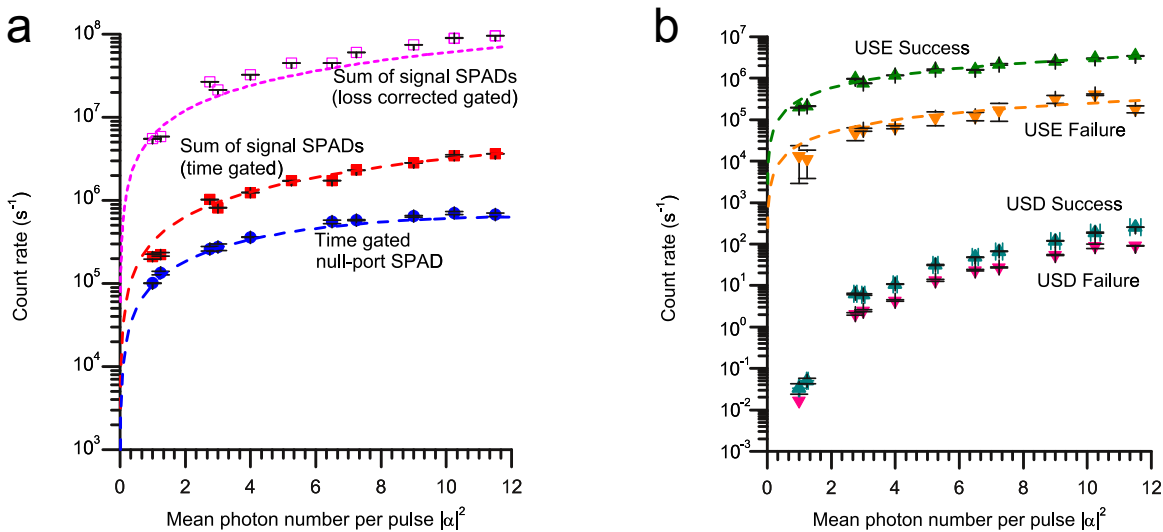}
\caption{ \label{Fig:results}
(a) Experimentally measured time-gated detector events for Charlie. The time-gated signal count-rate is the detector click rate summed over all four of Charlie's signal SPADs,
filtered by a window of  $\pm$1~ns around the expected pulse arrival time.
The  ``Loss corrected gated rate" is the calculated time-gated count rate at the signal output of the multiport.
(b) Rates of successful and failed measurements for one recipient. ``USE Success'' means that at least one state was correctly excluded using quantum state elimination. That is, the state Alice actually sent was not excluded, and at least one other state was excluded.  ``USE Failure" means that the state Alice actually sent was excluded.
``USD Success'' is the rate of successful unambiguous state discrimination (the correct state was obtained) while ``USD Failure'' is the rate of unsuccessful unambiguous state discrimination (the obtained state differs from that sent by Alice).
Data-points represent experimental results and dashed lines are theoretical predictions~\cite{RobustGHz}.  Experimental data are averaged over several measurements and
error bars in the count rate
are the standard deviation.  Horizontal error bars for the mean photon number are dominated by a worst-case assumption that the pulse-to-pulse variation in the output power of our laser is the experimentally measured maximum of $\pm$1.5\%.}
\end{figure*}

To prevent repudiation, recipients must ensure that they are sufficiently unlikely to disagree on the validity of a message. Here, as in \cite{OurNatComm}, this is achieved using an all-optical fiber multiport, shown in Fig.~\ref{Fig:setup} within dashed brown lines. Bob and Charlie split the pulses received from Alice using a 50:50 beam splitter. Bob sends to Charlie half of the pulse he received from Alice, and Charlie does correspondingly. Bob then combines the component he received directly from Alice with the component he received from Charlie on another 50:50 beam splitter, and Charlie again does correspondingly. This symmetrizes Bob's and Charlie's quantum states for each position in the signature sequences, so that their measurement statistics at the output of the multiport are identical. By choosing a lower allowed fraction of mismatches $s_a$ for accepting a message received directly from Alice, and a higher allowed fraction $s_v$ for verifying that a forwarded message is genuine, it can be made unlikely that two recipients will disagree on the validity of a message, see~\cite{QDS, OurArXiv} and Appendix \ref{Supp}.

Moreover, the multiport guarantees that even if Alice uses general, possibly entangled quantum states, she still cannot make Bob and Charlie significantly disagree on the validity of a signature. In addition, by considering counts at the multiport null-ports, the recipients can guard against certain types of forging.

In \cite{OurArXiv} and in the Appendix \ref{Supp} we show that the probabilities for repudiation and forging decay exponentially in $L$, by suitable choice of protocol parameters depending on the properties of an actual implementation. The scheme can also be made {\it robust}, that is,
if all parties are honest, then the protocol runs as intended with high probability. In any implementation, errors will occur even if all parties are honest. Therefore to ensure robustness one should for example select $s_a>0$.

Defining the level of security in QDS is not straightforward, since
different parties may be honest and dishonest. Here we assume that one chooses values of $s_a$ and $s_v$ such that the probabilities for repudiation, forging, and for rejection if all participants are honest
are all equal (see Appendix \ref{Supp}).
The probability of any of these undesirable events occurring is then
\begin{equation}\label{failure_prob}
P(\textrm{failure})\leq \exp(-\frac{g^2}8 L),
\end{equation}
where $g$ is the gap giving a lower bound on the advantage that someone (e.g. Alice) has if she knows the signature, compared to someone else (e.g. a forger) who makes a guess
by performing a measurement on the signature copies (see Appendix \ref{Supp}). In this paper, we will call the failure probability the \emph{security level of the QDS scheme}. Eq. (\ref{failure_prob}) shows that a greater gap $g$ gives better protocol performance.

The figure of merit that we will use to quantify the performance of our experiment is the length of the signature $L$ required to sign a ``half-bit'' message for a given security level.
One can also define the \emph{rate} of the signature as the number of bits per second that can be signed securely, given the clock rate of the source used.
Our experiment uses a clock rate of 100~MHz,
due to the temporal response profile of the Geiger-mode silicon single-photon avalanche diodes (Si-SPADs)
\cite{GHzDetector}.

We explored amplitudes
from  $|\alpha|^2=1$ to
11. Coherent states are generated by a temperature-stabilized pulsed vertical cavity surface emitting laser (VCSEL)
with wavelength
850.17 nm,  attenuated to the desired mean photon number per pulse $|\alpha|^2$,

defined at the launch from Alice into the multiport.
For a given run of the experiment with some $|\alpha|^2$, we registered the phases that Bob and Charlie ruled out.

This experimental data gives the probabilities of excluding particular states, given that Alice sent a certain state.
All losses are included, since these probabilities are
determined from the experimentally measured ratio of detection events to the total number of pulses sent by Alice.
For the QDS scheme to be secure,
an honest participant must be able to detect a difference between forged and genuine signatures (see Appendix \ref{Supp}).
How large this difference is determines
$g$ (see \cite{Min-cost,Supp}), and therefore through Eq. (\ref{failure_prob}) determines the length $L$ required for a desired security level.
The gap $g$ is proportional to the the transmittance (one minus the losses) (Appendix \ref{Supp}) and therefore
the length $L$ for a fixed security level decreases quadratically as the transmittance increases.
In short, the difference between the success and failure probabilities for USE determines how well a participant can identify a false declaration.

Experimental results
are shown in Figure \ref{Fig:results}.
Each data point represents the mean of several measurements. Vertical error bars are the standard deviation, and horizontal error bars the uncertainty in the mean photon number due to a pulse-to-pulse variation of  $< \pm$1.5\% in VCSEL intensity.  In Figure \ref{Fig:results}b ``USE Success" means that at least one state was eliminated, as long as the state that Alice actually sent was not eliminated. ``USE Failure" means that the state that Alice actually sent was eliminated. The success probability for USD is also shown, and is considerably lower than
for USE.
With USE, one sometimes excludes more than one state. All ``USD Success" events are
included in the ``USE Success" data.
As already noted, with USE many events which would count as undetected in USD will now contribute to the detection of forging, in addition to all USD success events.
The difference between the success and failure probabilities is greater for USE than it is for USD, similarly indicating that USE leads to a greater chance of detecting forging.

For all investigated values of $|\alpha|^2$, the success probability for USE is much higher than the failure rate. For higher failure rates, one has to set acceptance and verification thresholds $s_a$ and $s_v$ higher to ensure robustness. This in turn increases the signature length required to ensure the same security level.
The primary cause of a ``failure", for both USE and USD, was the fringe visibility of the detection setups, which was 80.9\%.
The multiport has a fringe visibility of 99.7\%.

When determining the optimal
$|\alpha|^2$, one has to consider
that the gap also depends on $p_{\textrm{min}}$, which is the minimum error probability that a forger
obtains if he tries to guess Alice's signature by measuring a copy of the quantum signature (Appendix \ref{Supp}). For very small $\alpha$, $p_{\textrm{min}}$ is large, but detecting a false declaration is difficult,
while for very large $\alpha$, $p_{\textrm{min}}$ is small but detecting a false declaration is relatively easy.
Since the  ability to detect a false declaration is estimated from experimental data,
and does not have an analytical expression, it is not straightforward to determine the optimum $\alpha$.
In our experiment, the best gap $g=1.20\times10^{-6}$ occurs for $|\alpha|^2=1$. For a security level of $0.01\%$ this gives $L=5.10\times10^{13}$ to sign a ``half-bit''.
This is an impractical signature length, and below we will comment on planned improvements
in order to make this rate more practical.

The signature length $L$ increases with increased distance between parties, since $g$ in Eq. (1) is proportional to the transmittance $\eta$. For example, if  $\eta$ is squared,
then the $L$ required for the same level of security will increase by a factor of $\eta^{-2}$.
In any event, if honest recipients see a difference between a forged and a genuine signature, however small,
then it is always possible to find
values of $s_a,s_v$ and $L$ to give a desired level of security.

To conclude, we have experimentally demonstrated a first realization of a QDS scheme which does not require long-term quantum memory, and where the recipients use quantum state elimination.
This is an important step in developing practical QDS systems. Our experiment uses phase-encoded coherent states. Recently, Arrazola and L\"utkenhaus suggested using phase-encoded coherent states for quantum fingerprinting~\cite{Arrazola}.
In our demonstration, due to the difficulty of stabilizing a multiport with long optical paths the sender and receivers were only separated from each other by approximately five meters of optical fiber.
Separate reference signals are needed for calibration before signature transmission, and as phase reference for the USE measurements. Tampering with reference pulses by a malevolent party should not lead to higher probability of forging or repudiation than tampering with signal states themselves~(Appendix \ref{Supp}). Also, reference signals can be bright, and thus can in principle be fully monitored through quantum tomography.

We are currently exploring three
changes to significantly improve
performance. First, by extrapolating data from a recent experiment on USE, we expect the optimal
$|\alpha|^2$ to be around $0.5$. Due to the high losses of this early prototype we were unable to successfully resolve measurements at this $|\alpha|^2$. The second
improvement is to use a protocol that does not require a multiport, to decrease loss.
Non-repudiation then needs to be guaranteed in an alternative way, similar to our recently proposed alternative QDS schemes \cite{No_multiport}, which could be modified to use phase-encoded coherent states, similar to the current realization. We estimate that implementing these changes will result in a gap of $g=1.96\times10^{-4}$, and length $L=1.19\times10^{9}$ for a security level of $0.01\%$. This protocol also potentially offers increased distances between sender and receivers.

Finally, increasing the clock rate, and therefore the transmission rate, is possible.
The phase modulators, VCSEL, and driving electronics are capable of clock-rates up to 3.3~GHz. In the system described in this Letter we did not employ such clock-rates due to the limitations of the time-stamping electronics~\cite{Jitter}.

\acknowledgments

This work was supported by the UK Engineering and Physical Sciences Research Council (EPSRC) through grants EP/G009821/1, EP/K022717/1, and the EPSRC Doctoral Prize fellowship grant. P.W. gratefully acknowledges partial support from COST Action MP1006.


\appendix

\section{Supplementary Material}\label{Supp}

\subsection{Introduction}

In the first part of this Supplementary Material we give details of the experimental methods used. We then more formally outline the protocol for a one-bit message. For longer messages, the protocol can be iterated, and the security parameters readjusted to compensate for the repeated (but independent) use of the single bit protocol. We then describe the unambiguous state elimination (USE) measurement employed by the recipients, and discuss some crucial properties of the multiport which the security against repudiation relies on.  After this we state definitions of security, state some inequalities which are used in the security analysis, and then proceed to analyze security against repudiation and forging, and the robustness of the three-party no-memory QDS protocol, realized using
coherent states with four possible phases.

We conclude with a summary of the performance of the protocol, suggestions for modifications that would increase the performance of the protocol, and estimate of the effect of these improvements.

\subsection{Experimental Methods}

In the experiments detailed in the main letter,
the coherent states are generated by a temperature-stabilized pulsed vertical cavity surface emitting laser (VCSEL) emitting at a wavelength of  850.17 nm and attenuated to the desired mean photon number per pulse $|\alpha|^2$  using spatial interception by a knife edge connected to a computer controlled stepper motor.
The uncertainty in Alice's phase encoding was primarily due to amplitude fluctuation in the electrical driving signal to the lithium niobate (LiNbO$_3$) phase modulator and corresponded to $\pm$1.6 $\times$ 10$^{-3}$ radians, or $\pm$0.2\% of the separation between the four phases used.
The multiport exhibited a loss of 7.7 dB, the receiver's beamsplitter 5.1 dB and each demodulation interferometer 9.1 dB.

Photodetection is carried out by commercially available thick junction  Geiger-mode silicon single-photon avalanche diodes (Si-SPADs)~\cite{SPAD} with a mean detection efficiency of 40.5\% (at a wavelength of 850 nm), dark count rate of 320 counts per second, and timing jitter of 380 ps, and time stamping electronics that could record photon arrival times at time intervals of 1~ps, but exhibited 12~ps independent timing jitter~\cite{Jitter}. Although using an emission wavelength near 1300 $\mu$m or 1550 $\mu$m would permit compatibility with standard telecommunications optical fiber, detector technologies for these wavelengths~\cite{Photodet1} are not as advanced as those at visible and shorter near infra-red wavelengths, and suffer from higher dark count and afterpulsing rates~\cite{Photodet2} which would increase error rates.

Although the dark count-rate of the detectors is relatively small and the possibility of intersymbol interference low, we time-gate the raw detector events using a window of duration $\pm$1~ns centered on the expect photon arrival time to temporally filter spurious counts from the recorded events.  At a delay of 10~ns after the detection peak the probability of a photo-generated count was 2 $\times$ 10$^{-6}$~\cite{GHzDetector}.
The time-gated count rate is approximately 91\% of the raw count rate.

Phase is a relative measurement and a reference for zero phase must be provided in some manner.  In our system we employ a technique frequently employed in
quantum key distribution systems which use phase-encoded states~\cite{Marand}. An asymmetric unbalanced Mach-Zehnder interferometer is used to provide a time-delayed reference pulse with zero phase, which  propagates through the same fiber as the phase-modulated ``signal". The phase reference pulse is delayed by 5~ns relative to the signature state pulse at Alice, and the delay is canceled in each receiver. The phase reference and signature state pulses have orthogonal polarizations and therefore can be correctly routed by the receivers using a polarization beamsplitter (PBS)~\cite{Marand}. All fiber components are constructed from polarization-maintaining ``panda eye" fiber~\cite{Panda} with a core diameter of 4.4 $\mu$m. Polarization routing
increases the time-gated detector count-rate of our system from 43\% of the raw count-rate as observed using a 50:50 beamsplitter to 91\% of the raw count-rate when using a PBS.
We comment further on establishing a common phase reference below in Section IV, in connection with the theoretical description of the USE measurement.

Changes in temperature or mechanically induced stress will result in changes in the relative path-lengths of the interferometers
in the system. These changes in path-length lead to changes in the visibility and, typically, to increased errors.  The recipients employ variable delay air-gaps consisting of a fixed collimating launch lens and a collection lens connected to a piezo-electric computer-controlled linear actuator to maximize the fringe visibility in the multiport and the demodulation interferometers~\cite{RobustGHz, GHzDetector}. The piezo-electric linear actuators had a step size of approximately 15 nm and a maximum travel of approximately 1.5 $\mu$m. An adjustable position knife edge was placed in the collimated beam in each air-gap to match the loss in each optical path. A mutually known tuning signal is used to establish maximum fringe visibility prior to signature transmission.

For this demonstration a shared 10~MHz Rb clock was used to provide a reference for both sender and receivers, but it is possible to use separate clocks with connections to the signals broadcast by the global positioning system (GPS) satellites to maintain a shared time reference~\cite{GPS}.

\subsection{Protocol outline}

\begin{itemize}
 \item[] \textit{Distribution stage}
 \begin{enumerate}
\item For each possible future message $k=0,1$, Alice generates
two copies of a sequence of coherent states (called \emph{quantum signatures})
$QuantSig_k = \bigotimes_{l=1}^L \rho_l^k$
where
$\rho_l^k = \dm{b_l^k \alpha}$, $\alpha$ is a real positive amplitude, $b_l^k \in \{1,i,-1,-i \}$  are randomly chosen, and $L$ is a suitably chosen integer.
The state $QuantSig_k$ and the sequence of signs $PrivKey_k = (b_1^k, \ldots b_L^k)$ are called the \textit{quantum signature} and the \textit{private key}, respectively, for message $k$.
The individual state $\rho_{l}^k$ we call
the $l^{th}$ \emph{quantum signature element} state for
message $k$.

\item Alice sends one copy of $QuantSig_k$
to Bob and one to Charlie,
for each possible message
$k=0$ and $k=1$.
\item \label{prot:USD}
Bob and Charlie send their
sequences $QuantSig_k$ for
$k=0$ and $k=1$, one signature element at a time,
through the QDS multiport.
For each signature element they (a) note whether photons are registered at their multiport null-port.
They also
(b) measure the multiport signal states
using the USE measurement for $\left\{ \ket{\alpha}, \ket{i\alpha},\ket{-\alpha},\ket{-i\alpha}\right\}$ (see below). For every element of each quantum signature (for $k=0,1$), they store which detectors detected photons;
each detector rules out one possible phase state.
They therefore store sets of six numbers (hexaplets) of the form $\{k,l', a_0^{k,l'},a_{\pi/2}^{k,l'},a_{\pi}^{k,l'},a_{3\pi/2}^{k,l'}\}$, where $1\le l'\le L$ and $a_\phi^{k,l'}\in\{0,1\}$. Here $a_\phi=0$ means that no photons were detected at the $\neg\phi$ detector (that is, the phase $\phi$ is not ruled out), while $a_\phi=1$ means that there was at least one photon detected at the $\neg\phi$ detector (that is, the phase $\phi$ is ruled out for this element). Note that by $\neg\phi$ detector we symbolize the ``not $\phi$'' detector.

 \end{enumerate}

\item[] \textit{Messaging stage}
\begin{enumerate}

\item To send a signed one-bit message $m$, Alice sends $(m,PrivKey_m)$ to the desired recipient (say Bob).

\item
Bob checks whether $(m,PrivKey_m)$ matches
his
stored sequence. In particular, he counts the number of elements of $PrivKey_m$ which disagree with his stored hexaplets. Therefore, for a given element $l$ of the signature, if Alice's declaration was $\phi$, Bob needs to check if $a_\phi$ is 0 or 1. If $a_\phi=1$, he registers one mismatch. In other words, a mismatch is registered whenever Alice's declaration for a given element has been eliminated by Bob's USE measurement. Bob checks whether the number of mismatches is below $s_a L$, where $s_a$ is an \textbf{authentication threshold}.

\item Provided the authentication threshold was not breached, before accepting the message, Bob checks that he has no reason to abort the protocol. If the number of signature elements for which non-zero null-port counts are registered
is higher than a threshold $r L$ for some $0\leq r <1$ he aborts. If the authentication threshold was not breached, and the protocol has not been aborted, Bob accepts the message coming from Alice.

\item To forward the message to Charlie, Bob forwards to
Charlie the pair  $(m, PrivKey_m)$ he received from Alice. Charlie tests for mismatches similarly to Bob, but to protect against repudiation by Alice, he uses a different threshold. Charlie checks if the number of mismatches  is below $s_v L$ where $s_v$ is the \textbf{verification threshold}, with $0 \leq s_a < s_v <1$.

\item For Charlie to accept the forwarded message, provided the verification threshold was not breached, he
confirms that he has no reason to abort the protocol by checking the null-port counts in the same way as Bob.

\end{enumerate}
\end{itemize}

\subsection{The USE measurement}

In quantum communication it is important to be able to distinguish different signal states from each other. Signal states may be non-orthogonal, either on purpose, for example in protocols for quantum key distribution, or because of unavoidable noise.
Quantum measurements aiming to distinguish between quantum states can be optimized in different ways, for example by minimizing the probability to select the wrong state (minimum-error measurement), by minimizing the average cost of selecting a measurement result (minimum-cost measurement), by maximizing the classical information in the result about what state was sent (maximum mutual information measurement) or by requiring that an obtained result is always correct, but the measurement is allowed to sometimes give an inconclusive answer. The last type of measurement is known as unambiguous state discrimination (USD). In all cases, these measurements aim to determine what state was sent, and are therefore called ``quantum state discrimination'' measurements.

One can also instead consider ``quantum state elimination'' or ``quantum state exclusion", which in a sense is a generalization of state discrimination \cite{stevebook,OppenUSE,CavesPRA}.
The difference is that one no longer aims to identify which state was sent, but instead which state(s) were \emph{not} sent, i.e. to rule out or eliminate one or more of the possible states. If one eliminates $N-1$ states, this is the same as identifying the state that was sent, and it is in this sense that state discrimination is a special case of state elimination. Just as for quantum state discrimination, one can for example aim for minimum-error state elimination, minimum-cost state elimination or unambiguous state elimination (USE). In~\cite{CavesPRA}, Caves \emph{et al.} discussed the compatibility of quantum-state assignments, and this can be seen as quantum state elimination.

The case of excluding one of three states was also explicitly given. In~\cite{stevebook}, Barnett discussed unambiguous quantum state elimination of a single state out of many possible ones, and was the first to use the term. More recently, this type of measurement was termed ``quantum state exclusion" in~\cite{OppenUSE}.

In our protocol, Bob and Charlie perform unambiguous quantum state elimination. The advantage of using unambiguous measurements
is that whenever Bob or Charlie obtains a result other than an inconclusive result, they can be certain that this result is correct (in the ideal case\footnote{In an experimental realization, measurement results are no longer error-free. However, the probability of a correct outcome, conditioned on obtaining an unambiguous outcome, will still be higher than for other types of optimal quantum measurements.}). Bob does not know for which of the elements of the signature Charlie has obtained unambiguous outcomes (or what the outcomes are). In order to forge, he needs to make a guess for \emph{all} the elements of the signature. Bob's best forging strategy will therefore be some type of minimum-cost state discrimination measurement.

The advantage of using state {\it elimination} instead of state {\it discrimination} is that the success probability for eliminating fewer than $N-1$ out of $N$ states can be (much) higher than the success probability of full USD, especially when considering experimental imperfections. Also, even if one aims to perform a full USD measurement, when it fails to unambiguously identify the state it may sometimes still rule out some of the states. This is the case for the best known linear optical realization of a USD measurement among symmetric coherent states, suggested by \cite{vanEnk} and realized for four coherent states in~\cite{ExpUSD}.
For an application such as quantum digital signatures, it is advantageous for Bob and Charlie to use all available information of what states have been ruled out. This is what led us to employ
a USE rather than USD measurement in our realization of the QDS protocol.

\begin{figure}
 \includegraphics[height= 6 cm]{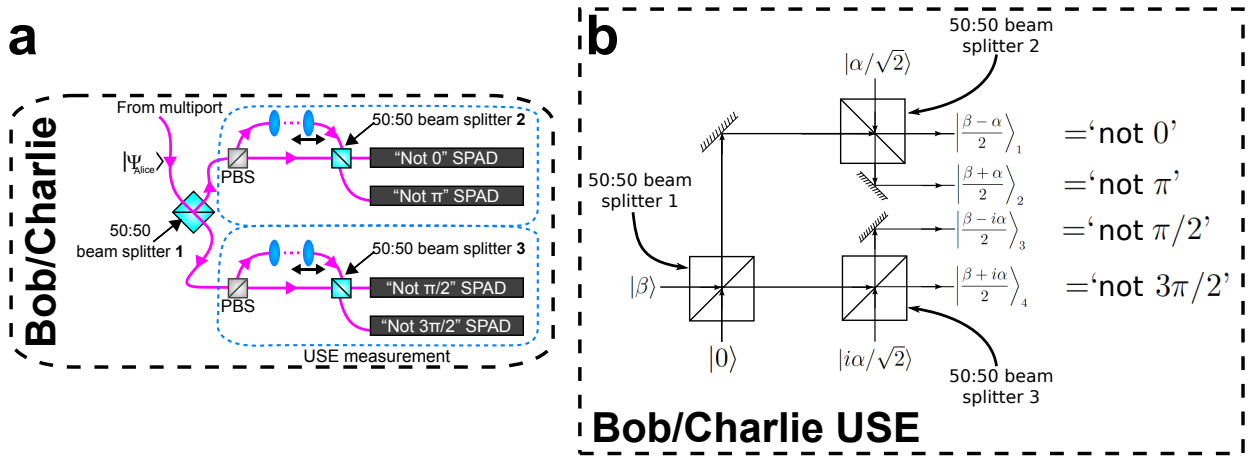}
\caption{\label{Fig:USE} Figure (1a) The part of Figure 1 in the main paper which shows the detection system for unambiguous state elimination with four coherent states.  PBS denotes a polarization beam splitter used to separate and route the orthogonally polarized signal and reference pulses.  The signal pulse takes the long curved path, entering in the top input ports of beam splitters 2 and 3, while the reference takes the short straight path, entering in the left-hand input ports of beam splitters 2 and 3.  The polarization of the signal pulse is rotated by 90 degrees in the long curved path so that it matches that of the reference pulse.  The relative difference in length between the signal and reference paths is chosen to be equal to that introduced by Alice during state preparation, so that the two pulses arrive at the final beamsplitters (labelled 2 and 3) at the same time and experience interference \cite{Townsend1993}.
Figure (1b) The setup for unambiguous state elimination for four coherent states, using the same beam splitter labels as in part (a). For clarity, this and other subsequent figures in the Supplementary Material are drawn indicating free-space optics with beam splitter cubes and mirrors, whereas our  experiment uses optical fiber as shown in Figure 1 of the main paper and part (a) of this figure. The beam is first directed onto the 50:50 beam splitter 1. One of the resulting beams is interfered with a reference beam in the state $|\alpha/\sqrt{2}\rangle$ using beam splitter 2, and the other one is interfered with a reference beam in the state $|i\alpha/\sqrt{2}\rangle$ using beam splitter 3. The reference beam is obtained from the reference pulse that Alice sends, as can be seen in part (a) of this figure. The four output states correspond to the four elimination outcomes.}
 \end{figure}

In the current QDS protocol, Alice chooses from four
coherent states $\{\ket{\alpha},\ket{i\alpha},\ket{-\alpha},\ket{-i\alpha}\}$. The USE measurement
results in eliminating one, two or three
possible states (unambiguous or conclusive result) or none at all (ambiguous or inconclusive result). For each signature element,
Bob directs the beam received from Alice onto a 50/50 beam splitter, and Charlie does the same. One of the resulting beams is interfered on another 50/50 beam splitter with a reference beam in the state $|\alpha/\sqrt{2}\rangle$, and the other one is interfered with a reference beam in the state $|i\alpha/\sqrt{2}\rangle$, as shown in Figure~\ref{Fig:USE}.
Note that the reference beam is obtained from a reference pulse sent by Alice, in order to
ensure that the outgoing states are as expected on recombination at beam splitters 2 and 3 (see below). Here we assume that the reference pulse is not tampered with. However, it is worth noting that intuitively one cannot obtain further repudiation and forging possibilities through tampering with the reference pulse. Repudiation is guaranteed by the symmetrization of the states due to the multiport, and this does not depend on the timing of the reference pulse. As far as forging is concerned, tampering with the reference pulse cannot help Bob to obtain further information about the signature, and since the original phase sent by Alice is unknown to him, he cannot use the reference pulse to steer the outcome of Charlie's USE measurement towards his declaration either.
In principle, the correct phase reference states for the USE measurement could also be obtained without a reference pulse sent from Alice to Bob and Charlie. If the parties instead have synchronized clocks, then Bob and Charlie could themselves prepare the required USE reference states.

Given the above considerations, we can now see that the resulting output state, assuming that the incident state was $|\beta\rangle$, and assuming no losses, is
\begin{equation}
|\psi_{out}\rangle=\ket{\frac{\beta-\alpha}2}_1\otimes\ket{\frac{\beta+\alpha}2}_2\otimes\ket{\frac{\beta-i\alpha}2}_3\otimes\ket{\frac{\beta+i\alpha}2}_4,
\end{equation}
where the four output modes refer to either Bob's or Charlie's output modes.
From this it is clear that if the state that Alice sent was $|\alpha\rangle$ to start with, then the first mode will contain the vacuum state. Therefore, if any photons are detected in this mode, it follows that the state Alice sent cannot have been $|\alpha\rangle$. Similarly, if Alice sent the state $|i\alpha\rangle$, then the second mode necessarily contains the vacuum state, and therefore detecting any photons here unambiguously rules out the state $|i\alpha\rangle$. Seeing any photons in output modes 3 and 4 rules out the states $|-\alpha\rangle$ and $|-i\alpha\rangle$, respectively.

To summarize, any detector clicking rules out one possible state. If three detectors click we have fully and unambiguously (in the ideal case) determined which state Alice sent. In other cases, Bob or Charlie may only have eliminated some of the possible states.
For every signature element, Bob and Charlie register which detectors (if any) click. In the messaging phase, when Alice (or a forger) declares what a particular signature element state was, Bob and Charlie check if this state was ruled out by their elimination measurement. If it was, they register a mismatch.
Losses in the setup will not affect the conclusive nature of the outcomes, only the success probability. Dark counts and other imperfections, on the other hand, may lead to errors in conclusive outcomes. This will be discussed later in connection with experimental results.

We should also note that the USE measurement described above may not be the optimum USE measurement. Exploring the properties of optimal USE measurements is an interesting problem. Nevertheless, the great advantage is that the realization we have described requires only linear optics and commercially available photodetectors.

\subsection{The multiport}

\begin{figure}
 \includegraphics[width=10cm,height= 10 cm]{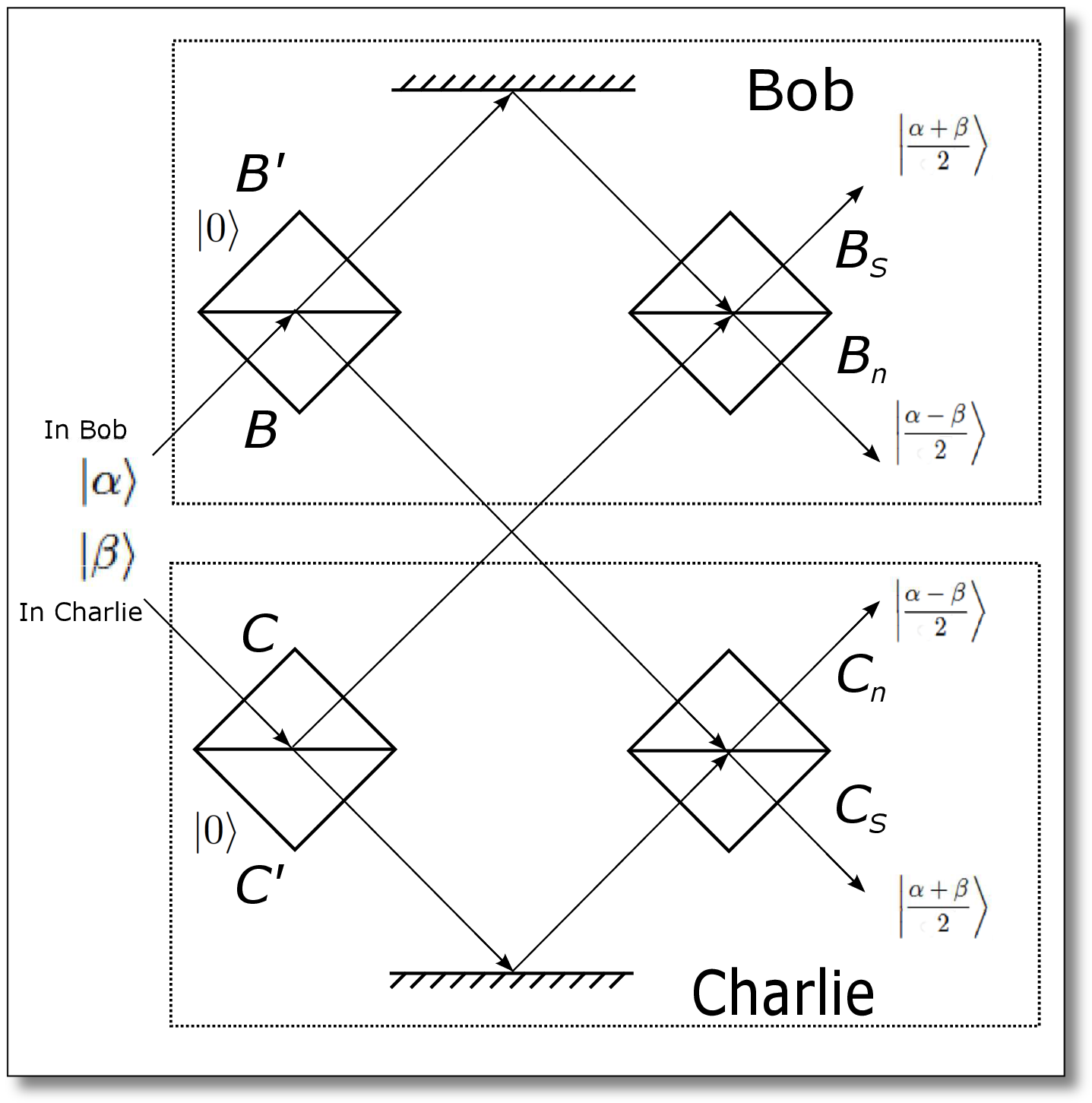}
\caption{\label{Fig:mult1} The multiport consists of four beam splitters. The top two belong to Bob and the bottom two to Charlie. In the figure above, output coherent states are given for coherent states $|\alpha\rangle$ and $|\beta\rangle$ in the non-empty input modes. Note that here we drew a free-space version, while in the experimental implementation we had a fiber-based version of the multiport. }
 \end{figure}

The multiport is a passive linear optical device with four input modes and four output modes, comprising four
50/50 beamsplitters.
Two of the input modes always contain the vacuum state.
The  beam splitters act on the field operators according to
\EQ{
\binom{\hat{a}_{out}^{\dagger}}{\hat{b}_{out}^{\dagger}} = \dfrac{1}{\sqrt{2}} \left(
\begin{tabular}{cc}
1 & 1\\
1 & -1\\
\end{tabular}
 \right)
\binom{\hat{a}_{in}^{\dagger}}{\hat{b}_{in}^{\dagger}}.
}
 The
 multiport and the input-output relations for coherent state inputs are illustrated
 in Figure \ref{Fig:mult1}.
The top two beam splitters are held by Bob and the lower two by Charlie.
Since two of the four input modes of the multiport are always set to the vacuum state by Bob and Charlie, we will refer to the remaining two modes as the input modes of the multiport.
For an input state where two of the input modes are in the vacuum state while the other two are coherent states $\alpha$ and $\beta$,
the multiport acts according to
\EQ{\ket{\alpha}_B\otimes\ket{\beta}_C\otimes\ket{0}_{B'}\otimes\ket{0}_{C'}\rightarrow \ket{\frac{\alpha+\beta}2}_{B_s}\otimes\ket{\frac{\alpha+\beta}2}_{C_s}\otimes\ket{\frac{\alpha-\beta}2}_{B_n}\otimes\ket{\frac{\alpha-\beta}2}_{C_n}.}
Alice controls the (non-vacuum) input states of Bob and Charlie.
As proven in detail in~\cite{QDSPRL}, it follows that the reduced state in Bob's signal port and the reduced state in Charlie's signal port are identical, for each individual signature element, even if the initial state sent by Alice was a general, possibly entangled state. That is, the reduced states
\EQ{\rho_{B_s}=\rho_{C_s}}
for each signature element. This holds even if Alice employs any type of entanglement, such as entanglement between different signature elements, entanglement between Bob's and Charlie's states, or entanglement with some other state which could be retained by Alice. One can further prove a somewhat stronger property, that the resulting state for Bob and Charlie, for one signature element position, is symmetric under swap of Bob's and Charlie's subsystems.

The multiport we consider here has two recipients, but it can also be generalized to many recipients~\cite{ErikaOrig}.

\subsection{Definitions of security}

The presented Quantum Digital Signature protocol is designed to be immune to two types of malicious activities: forging and repudiation.  Immunity to forging means that any receiving party will with high probability reject any message which was not sent by Alice herself.  Immunity to repudiation guarantees that if Bob accepts a message from Alice, then with high probability
the same message will pass verification with Charlie as well. That is, Alice cannot make Bob and Charlie disagree on the validity
(and consequently the content) of her message. More formally we have the following:

\begin{itemize}

\item We say that a protocol realizing QDS is \textbf{secure against forging} if the probability of a recipient successfully producing, without receiving it from Alice, a private key for a message $m$, which will pass verification by the other recipients, is \emph{decaying exponentially quickly as a function of the quantum signature length} $L$.

\item We say that a protocol realizing QDS is \textbf{secure against repudiation} if, for any malicious activity by Alice, the probability of a message failing verification with one recipient once it has already passed authentication with another recipient is \emph{decaying exponentially quickly as a function of the quantum signature length} $L$.

\item We say that a protocol realizing QDS is \textbf{robust} if, when all parties are honest, a message will be authenticated and verified except with a probability \emph{decaying exponentially quickly as a function of the quantum signature length} $L$.

\end{itemize}

We will examine one isolated run of the protocol.
A more general treatment, such as analysis of composable security, we leave for future work.

Defining the level of security for a QDS scheme is complicated. Unlike for QKD, it is not fixed which participants are honest and which are malevolent. In other words, some choices of parameters may lead to a certain level of guaranteed security against forging and a different level for security against repudiation. However, one can

choose $s_a$ and $s_v$ such that the resulting probabilities for any undesirable event (repudiation, forging or honest rejection) are equal. In this case, we will call the probability for an undesirable event the \emph{security level} of the QDS scheme. In the last section, we will comment further on this, and on the resulting rate and length $L$ required to sign a half-bit message for a given security level.

Both for repudiation and for forging one can distinguish different types of malicious attacks. In {\it individual} attacks, the cheating party employs a strategy separately and independently for each signature element. In {\it collective} attacks, there may be classical correlations between strategies for different signature elements. These attacks can also be called {\it separable}. {\it Coherent} attacks are the most general type; here a cheating party can employ any type of correlations, including entanglement and measurements in an entangled basis. For forging, we can additionally distinguish between {\it passive}  and {\it active} attacks, depending on whether a forger acts maliciously only during the messaging stage, or also during the distribution stage. Security is proven for
all types of repudiation attacks, and all types of forging except coherent forging attacks. We will treat individual repudiation and
forging in more detail. The analysis for other types of attacks
follows the treatment in~\cite{QDSPRL}, and we will only point out the differences.
Intuitively, the protocol is secure also against coherent forging, but a full analysis of this remains an important task and is the subject of ongoing work.

\subsection{Hoeffding's inequalities}

In the security analysis, we will use the following forms of Hoeffding's inequalities \cite{hoeff}.

\LE Let  $X_1,\cdots, X_L$ be independent random variables, each attaining values $0$ or $1$. Let $\bar{X}=1/L\sum X_i$ be the empirical mean of the variables, and let $E(\bar{X})$ be the expected value of $\bar{X}$. Then for all $t\geq 0 $ we have

\EQ{P(\bar{X}-E(\bar{X})\geq t)\leq \exp (-2t^2 L)\label{Hoeffding 1}}

\EQ{P(E(\bar{X})-\bar{X}\geq t)\leq \exp (-2t^2 L)\label{Hoeffding 2}}

\EQ{P(|\bar{X}-E(\bar{X})|\geq t)\leq 2 \exp (-2t^2 L)\label{Hoeffding 3}.}
\EL

\subsection{Security against repudiation -- cheating Alice}

In a repudiation event, Alice manipulates the quantum signature states\footnote{We assume that all the classical information which is sent from Alice is cross-checked by Charlie and Bob over authenticated classical channels, so repudiation by tampering with classical messages is not possible.} which are sent in the distribution stage, in such a way that during the messaging stage, the same declaration will be confirmed by one party (checking against the threshold $s_a$) but rejected when forwarded to another party (who checks against the threshold $s_v$).
Repudiation is successful, if (say) Bob authenticates ($BA$), Charlie rejects ($CR$) and there is no abort ($NA$). To bound this probability, we can consider the minimum of the following two probabilities, Bob authenticating, and Charlie rejecting to verify, i.e.
\EQ{P(rep)=P(BA,CR,NA)\leq P(BA,CR)\leq \min \{P(BA),P(CR)\}\label{repudiation 1}.}
There are two things to note to understand intuitively why such a bound would be useful. First, we should note that due to the symmetricity of the multiport $P(BA)$ and $P(CR)$ are \emph{not} independent. Second, we should note that the threshold for Bob authenticating is lower than that of Charlie verifying.

Here we will
consider individual strategies by Alice.
General separable strategies and general coherent strategies, follow
by reduction to individual strategies as in~\cite{QDSPRL}. Therefore the best general strategy for Alice is given by individual attacks.

\subsubsection{Security against individual repudiation}

In this type of attack Alice sends quantum signatures, possibly different, to Bob and Charlie. Each pair of Bob's and Charlie's corresponding quantum signature elements are not correlated with other pairs (classically or through entanglement), but are not necessarily identical, pair to pair. That is, the global state of the quantum signature is in a factorized form, with respect to the partition in different signature element pairs. There may be entanglement within each Bob-Charlie signature element pair.
Because Bob and Charlie pass their quantum signatures through the multiport, the outbound
quantum signature element  at the signal ports of Bob and Charlie is always symmetric under swap of  Bob and Charlie. That is, the quantum signatures Bob and Charlie collect at their multiport signal ports are such that the reduced density matrix for each element of the signature of Bob is the same as Charlie's.

 For each signature element, Alice's declaration agrees with Bob's (Charlie's) measured eliminations with probability $p^i_1$ if the declaration is not eliminated, and disagrees with probability $p^i_{-1}$ if it is eliminated (mismatch). Those probabilities,
provided the multiport has been appropriately adjusted and calibrated, are identical for Bob and Charlie. We can define the \emph{average} probabilities $\bar{p}_j=1/L\sum_i p^i_j$, with $j\in\{1,-1\}$. Furthermore, we define as $\bar{X}_{-1}$ the empirical (observed) percentage of mismatches over the total number ($L$) of elements of the signature.

It is easy to see that Bob authenticating means that $\bar{X}_{-1}\leq s_a$, which using Eq. (\ref{Hoeffding 2}) leads to
\EQ{P(BA)=P(\bar{p}_{-1}-\bar{X}_{-1}\geq \bar{p}_{-1}-s_a)\leq \exp (-2(\bar{p}_{-1}-s_a)^2 L)\label{BA},}
provided that $\bar{p}_{-1}\geq s_a+\epsilon'$. Note that $\epsilon'$ is an arbitrarily small positive number. We note that if this is satisfied, the probability decays exponentially.

Charlie failing to verify means that $\bar{X}_{-1}\geq s_v$ which using Eq. (\ref{Hoeffding 1}) leads to
\EQ{P(CR)=P(\bar{X}_{-1}-\bar{p}_{-1}\geq s_v-\bar{p}_{-1})\leq \exp (-2(s_v-\bar{p}_{-1})^2 L)\label{CR}}
provided that $\bar{p}_{-1}\leq s_v- \epsilon'$. Alice's only freedom, in her attempt to repudiate, is to choose different $p^i_j$'s. In reality, Alice does not have full freedom for these choices (as the POVM elements of Bob and Charlie's measurements do not have orthogonal supports), but since we are looking for a bound, we may assume a worst-case setting and assume that she does have full freedom. Noting that $\bar{p}_{-1}+\bar{p}_1=1$, Alice can only choose the optimum for repudiation value of $\bar{p}_{-1}$. Therefore, provided that we choose
\EQ{s_v > s_a\label{Condition 1},}
the probability for repudiation decays exponentially for \emph{all} choices of $\bar{p}_{-1}$. From Eq. (\ref{repudiation 1}), we can see that Alice's best strategy to achieve repudiation is to choose $\bar{p}_{-1}$ in such a way that the minimum of BA and CR is the greatest. From Eq's (\ref{BA}-\ref{CR}) we get that the optimum value is for $\bar{p}_{-1}= (s_v+s_a)/2$. This gives the following bound for repudiation under individual attacks,
\EQ{P(rep)\leq \exp (-\frac{(s_v -s_a)^2}2 L)\label{Repudiation INID}.}
It is interesting to note that there is an optimal value of
 $\bar{p}_{-1}$ from Alice's point of view, and that her repudiation probability depends on this. Therefore, for the given bound we consider, it is no better for her to choose different values for different $p^i_{-1}$, and thus a strategy where Alice sends identical states for each signature element is at least as effective as full individual strategy for Alice.

\subsection{Security against forging}

As also noted in the main text, forging is defined as when a cheating party (say, Bob) convinces an honest party (Charlie) that Alice had sent a classical message $k$, when Alice has sent no message or another message.
To do so, the malevolent party needs to guess Alice's declaration. Throughout this section, when we mention declaration, and matching and mismatching with the declaration, this is the ``fake'' declaration of Bob, which is made after Bob uses all his resources to make the best possible
guess. Note also that security against forging as defined also ensures security against message tampering.

At the end of the distribution stage Bob/Charlie makes a USE measurement. The outcome of the measurement is that they rule out one or more of the possible phases. These outcomes should give a guaranteed minimum advantage to anyone who declares the actual phase sent, against anyone who declares some other phase. If this is the case, then there are choices of $s_a,s_v$ and $L$ that would make the protocol secure, as we will see below.

There are two different characterizations of forging attacks. The first concerns whether or not the malevolent party was dishonest from the beginning or only during the messaging stage. {\it Passive forging} is where Bob is honest during the distribution stage and simply attempts to guess the signature using his own copy, in such a way that his declaration gets accepted by Charlie. {\it Active forging} is where Bob is dishonest also during the distribution stage, where he attempts to modify Charlie's quantum signature so that it helps him subsequently during the messaging stage to make Charlie agree with his own declaration. Here we will first examine passive forging in greater detail,  and then argue that active forging can be reduced to modified passive forging attacks by placing a limit on the multiport null-port counts during the distribution stage.

The second characterization concerns the type of measurements and actions that the malevolent party can do, similar to as for repudiation. There are three types of attacks, (1) individual attacks, where Bob is restricted to individually measuring the elements of his signature and also to tampering with the response states individually, (2) collective attacks, where Bob is allowed to include classical correlations between the measurements and response states for different signature elements and (3) coherent attacks, where Bob is allowed to make any possible joint measurement and also send any possible entangled state to Charlie. In this paper, we will only examine the first two types of attacks and leave coherent forging strategies for future work.

From Eq. (\ref{Condition 1}) it is clear that it is easier to attempt to forge a message in the verification stage rather than in the authentication stage. In other words, it is easier to convince Charlie that Bob forwards the message received from Alice, than Charlie convincing Bob that he directly sends the message (pretending to be Alice).

We have defined as successful forging the scenario where Charlie verifies the fake message sent by Bob and there was no abort. Abort occurs only when the null-port of the multiport register above than $r L$. When Alice is honest, abort may thus occur only if Bob attempts to tamper with the signature at the distribution stage, that is, for active forging.

\subsubsection{Security against passive individual and collective forging}

In passive attacks Bob does not interfere during the distribution of the signatures, and therefore Charlie obtains as quantum signatures what Alice, who is honest, sends.
Bob's strategies comprise all the possible measurements he could perform on his copy of the quantum signature, resulting in a declaration for each signature element.
Bob aims to minimize the probability that his declarations will lead to a rejection, when compared with the eliminations that Charlie has obtained from his USE measurements.
Bob knows what measurement Charlie, who is honest, will make on the signature Charlie receives from Alice. In particular, Bob knows what Charlie's measurement statistics is, in what ways Charlie's measurement may be imperfect, and also what Charlie's verification threshold is, and Bob can adjust his measurement accordingly.
Bob's optimal strategy will be a measurement which maximizes the probability of Charlie accepting Bob's declaration as a whole. Recall, Charlie accepts if the number of incorrectly declared phases is below a certain number, determined by the verification threshold. The optimal measurement for this problem is a so-called minimum cost measurement~\cite{Helstrom}.
In particular, any declaration of Bob, as a whole, for all signature elements, has a cost equal to the probability that Charlie will reject
this declaration. This means that Bob's goal is {\it not} in general to minimize the probability of a mismatch for each individual element. Bob' goal is instead to meet the verification threshold. To take a simple example, suppose that there are two signature elements, and that Charlie will accept the message if there is at most one mismatch. Bob should then concentrate on avoiding two mismatches, and declarations with zero and one mismatches both carry no cost.

If Bob is limited to collective forging strategies, then it is in Bob's favor to minimize the probability of a mismatch for each individual element, as we will now argue. That is, it is optimal to make independent minimum-cost measurements for each signature element.
Collective strategies would allow Bob to change the measurement he makes on subsequent signature elements, conditioned on the outcomes he obtained for previous elements.
There is however no correlation between the phases of different signature elements, as Alice chooses these independently of each other. Charlie's eliminations also occur independently for each element of the signature. As long as Bob is not allowed to make a measurement in an entangled basis, any knowledge of the phase of one element does not alter his chances of guessing another element correctly, and does not alter what measurement he should make on this signature element. Therefore his best strategy is to make a minimum-cost measurement for each individual signature element.
In other words, Bob cannot benefit from classical correlations between his measurements for different signature elements.
Therefore, provided that one restricts attention to collective attacks, Bob's best strategy is to make a minimum-cost measurement for each signature element. We proceed to bound the probability of success for this type of individual forging attack.
We will prove that the forging probability is less than the minimum-error probability multiplied by the guaranteed advantage that an honest participant has.

For a given individual signature element, we define the cost matrix as a matrix where the rows corresponds to which state Alice sent ($\ket{\exp(i\theta)\alpha}$), while the columns correspond to the detectors $D(\neg\theta)$. Each matrix element $C_{i,j}$ can be taken equal to the probability that if the $i$'th state is sent, then Charlie's $j$'th detector clicks. This is because Bob should avoid declaring a phase that Charlie has eliminated. His cost for making a particular declaration will therefore be proportional to the probability that Charlie has ruled out this state. As we have mentioned earlier, in the ideal case, Charlie should never rule out the phase that Alice sends, and thus in the ideal case, Bob's cost matrix would have zeroes on the diagonal. However, due to losses and noise (mainly due to the multiport), we do have the following actual experimental cost matrix, for the case where $|\alpha|^2=1$,
\EQ{\label{exp cost matrix} C=\left( \begin{array}{cccc}
9.80\times10^{-5}, & 1.63\times10^{-4}, & 1.71\times10^{-4}, & 1.40\times10^{-4}\\
6.75\times10^{-5}, & 2.37\times10^{-5}, & 1.57\times10^{-4}, & 2.62\times10^{-4}\\
2.19\times10^{-4}, & 1.69\times10^{-4}, & 1.98\times10^{-5}, & 1.01\times10^{-4}\\
2.08\times10^{-4}, & 2.82\times10^{-3}, & 3.85\times10^{-5}, & 2.55\times10^{-5}\end{array} \right).}
In the experiment, we considered several different values for $|\alpha|^2$ ranging form $|\alpha|^2=1$ to $11$. We chose to give the cost matrix for $|\alpha|^2=1$ since, using the bounds we have, this gives the best performance (requires a smaller length $L$ for given security level) among the tested values of $\alpha$.
The minimum guaranteed advantage is the smallest difference between a diagonal element and an off-diagonal element in the same row. The reason for this will become apparent soon. For our case the minimum advantage is $1.30\times 10^{-5}$, as is seen from the fourth row.

In the general case, one has to assume that Bob has full knowledge of Charlie's actual measurement statistics, and thus this type of cost matrix should be used in a security analysis of an actual realization.
We recall that $\bar{X}_{-1}$ is the empirically observed percentage of mismatches. Bob succeeds in forging, if Charlie, after receiving Bob's (fake) declaration, finds less than $s_v L$ mismatches ($\bar{X}_{-1}\leq s_v$), and thus verifies. Consequently, Bob tries to guess the correct signature by making a minimum-cost measurement for each element signature, i.e. the measurement that, given the cost matrix in eq. (\ref{exp cost matrix}), minimizes the overall cost given by
\EQ{C_{min}=\min_{\{\Pi\}}\bar C(\Pi)=\min_{\{\Pi\}}\sum_{i,j}\eta_i C_{i,j}Tr(\Pi_j\rho_i).}
Here $\eta_i=1/4$ is the prior probability of each state (Alice sends each of the for phases with equal probability), and $\Pi_i$ are the elements of Bob's measurement, a probability operator measure (POM), also called a positive operator-valued measure (POVM).
 The minimum cost $C_{min}$ is the minimum possible probability that Bob's declaration for a single signature element has been eliminated by Charlie. We will give a bound for this cost, for our actual experimental realization, below. For now, we note that for each signature element we have a binary random variable that takes value ``$1$'' (match) or ``$-1$'' (mismatch), and $C_{min}$ is the expected probability for a mismatch. Since we assume that each of the elements is measured separately, the random variables are independent and the Hoeffding inequalities hold. Using eq. (\ref{Hoeffding 2}) the probability of forging (which occurs if the observed mismatches are fewer than the threshold $\bar X_{-1}L\leq s_vL$) is bounded as
\EQ{P(for)\leq P\left(C_{min}-\bar{X}_{-1}\geq C_{min}-s_v\right)\leq \exp \left(-2 \left(C_{min}-s_v\right)^2 L\right),}
provided that we choose $s_v$ such that $C_{min}\geq s_v$. If this holds, we see that the forging probability decays exponentially in the signature length $L$.

Before proceeding to derive a bound on the possible $C_{min}$ that Bob can achieve, we should point out that it is crucial that there is a gap between the minimum probability of a forger to have his declaration eliminated, $C_{min}$, and the probability $p_h$ of the correct declaration being eliminated if all parties are honest. Defining this gap as $g=C_{min}-p_h$, the choices of $s_a$ and $s_v$ for a robust and secure protocol should be within this gap (we elaborate more on parameter choices at the end).

We proceed to bounding the minimum cost of the measurement that a malevolent Bob performs. Methods for bounding the minimum cost and properties of the minimum-cost measurements that are used here can be found in \cite{Min-cost}.
Since we trying to find a bound on the forging probability, we assume the best case scenario for Bob, and therefore we are looking for a \emph{lower} bound on the minimum cost. We will use three properties of minimum-cost measurements. To start with, the minimum cost for a given cost matrix $C_{i,j}$ can always be bounded by the cost of any other strictly smaller cost matrix $C'_{ij}\leq C_{ij}$ for all $i,j$. Secondly, define a \emph{constant-row} matrix as a matrix of the form
\EQ{D=\left( \begin{array}{cccc}
d_1, & d_1, & d_1, & d_1\\
d_2, & d_2, & d_2, & d_2\\
d_3, & d_3, & d_3, & d_3\\
d_4, & d_4, & d_4, & d_4\end{array} \right).}
If we add or subtract such a constant-row matrix from any given cost matrix, then the measurement that minimizes the cost for the new cost matrix does not change, while the minimum cost simply shifts by a constant \cite{Min-cost}.
It is easy to see that the average
cost for any measurement, not just an optimal one, shifts by $\sum\eta_id_i$.
Intuitively, for a constant-row cost matrix, no matter what measurement one makes, the cost of all outcomes (i.e. the matrix elements in a given row) are equal. The cost
does not depend on what outcome is obtained, but only on which state Alice sends,
which in its turn depends on the prior probabilities $\eta_i$. The third property that we will use is the fact that the minimum-cost measurement for a cost matrix of \emph{error-type}, for the case of symmetric pure states, is the square root measurement (SRM). A cost matrix is of error-type if all the diagonal elements are zero, and all the off-diagonal elements are positive and equal with each other. We call this error-type, because for such a cost matrix,  guessing correctly has no cost (the diagonal elements), while any mistake carries an equal cost (the off-diagonal elements). The cost matrix for a minimum-error measurement, which results in  the minimum probability of error $p_{min}$, is of error-type, with off-diagonal elements having cost 1. This means that the minimum cost of any error-type cost matrix is proportional to the minimum-error probability $p_{min}$.

To bound $C_{min}$ for the cost matrix in eq. (\ref{exp cost matrix}), we first define $C^h_{i,j}=C_{i,i}$, a constant-row matrix for which all elements in every row are equal to the diagonal elements of the matrix $C_{i,j}$. Then we define a cost matrix $C'_{i,j}=C_{i,j}-C^h_{i,j}$, which has the same minimum-cost measurement as $C_{i,j}$, but with the minimum cost shifted by $C^h=1/4\sum_i C_{i,i}$. Finally we define another cost matrix $C^l_{i,j}$ which is strictly smaller than $C'_{i,j}\geq C^l_{i,j}$ for all $i,j$, such that $C^l_{i,j}=\min_{i\neq j} C'_{i,j}$ for $i\neq j$, and with zeroes on the diagonal. This final cost matrix $C^l_{i,j}$ is of error-type, and the corresponding minimum-cost measurement is therefore the SRM and the minimum cost is proportional to $p_{min}$. In our case, denoting the relevant minimum costs with $C^h$ and $C^l_{min}$, we have
\EQ{C_{min}\geq C^h+C^l_{min}.}
From eq. (\ref{exp cost matrix}) we obtain
\EQ{ C^h=\left( \begin{array}{cccc}
9.80\times10^{-5}, & 9.80\times10^{-5}, & 9.80\times10^{-5}, & 9.80\times10^{-5}\\
2.37\times10^{-5}, & 2.37\times10^{-5}, & 2.37\times10^{-5}, & 2.37\times10^{-5}\\
1.98\times10^{-5}, & 1.98\times10^{-5}, & 1.98\times10^{-5}, & 1.98\times10^{-5}\\
2.55\times10^{-5}, & 2.55\times10^{-5}, & 2.55\times10^{-5}, & 2.55\times10^{-5}\end{array} \right),}
\EQ{ C'=\left( \begin{array}{cccc}
0, & 0.65\times10^{-4}, & 0.73\times10^{-4}, & 0.42\times10^{-4}\\
4.38\times10^{-5}, & 0, & 1.33\times10^{-4}, & 2.38\times10^{-4}\\
1.99\times10^{-4}, & 1.49\times10^{-4}, & 0, & 0.81\times10^{-4}\\
1.83\times10^{-4}, & 2.57\times10^{-4}, & 1.30\times10^{-5}, & 0\end{array} \right),}
\EQ{ C^l=\left( \begin{array}{cccc}
0, & 1.30\times10^{-5}, & 1.30\times10^{-5}, & 1.30\times10^{-5}\\
1.30\times10^{-5}, & 0, & 1.30\times10^{-5}, & 1.30\times10^{-5}\\
1.30\times10^{-5}, & 1.30\times10^{-5}, & 0, & 1.30\times10^{-5}\\
1.30\times10^{-5}, & 1.30\times10^{-5}, & 1.30\times10^{-5}, & 0\end{array} \right),}
and also $C^h=p_h=4.18\times10^{-5}$. This is the cost for the honest scenario, in other words, the probability that Charlie has eliminated the state that Alice actually sent, so that there would be a mismatch even if all parties including Bob are honest, and Bob simply forwards Alice's correct declaration.
From the matrix $C^l$ we can see that the bound on the advantage of a correct declaration compared to a wrong declaration, i.e. the guaranteed advantage, is $guad=1.30\times10^{-5}$. The bound on the gap is then given by
\begin{equation}\label{gap-adv}g\geq C^l_{min}=p_{min}\times guad\end{equation}
We compute $p_{min}$ using the SRM \cite{Min-cost}, which for $\alpha^2=1$ gives $p_{min}=0.092$. We obtain $C^l_{min}=1.20\times10^{-6}$, which provides a bound on the gap $g$. To confirm that the bound on the minimum cost we obtained is relatively tight, we can obtain an upper bound for the minimum cost, along similar lines as the lower bound, which gives $C^u_{min}=2.19\times10^{-5}$. We therefore obtain
\EQ{6.37\times10^{-5}\geq C_{min}\geq 4.30\times10^{-5}.}
and
\EQ{2.19\times10^{-5}\geq g \geq 1.20\times10^{-6}.}

\subsubsection{Security against active individual and collective forging}

For active individual and collective forging attacks, the security analysis follows the treatment in~\cite{QDSPRL}.
We will discuss and give the expressions for
attacks where Bob behaves independently and identically for each quantum signature element (IID forging attacks).  However, independent, non-identically distributed strategies (INID attacks), and also collective forging strategies, reduce to, and are not more powerful than, IID forging strategies. Here we refer the reader to~\cite{QDSPRL}.

In essence, active forging can be limited in the following way. Both Bob and Charlie keep track of the number of detections at the null-port of the multiport. If one of the parties receiving the signature states (say Bob) is dishonest during the distribution phase, the state he sends may not be identical to the state that Alice sends, and if so there will on average be some detections at the null-port of the multiport. Since the protocol is aborted if the detections on the null-port exceed some number $r L$, we can bound how much
Bob can tamper with the signature states. This effectively reduces active forging to passive forging. Then the whole security analysis follows in a similar fashion as for passive forging, except that the verification threshold $s_v$ is effectively lowered by a function of $r$ which is the abort threshold. By choosing $r$ suitably small, we can make sure that the effective threshold $s'_v$ is still greater than $s_a$ so that the protocol remains exponentially secure.

\emph{IID attacks}. In active strategies, Bob is allowed to alter the quantum signature sent to Charlie during the distribution stage, through tampering with the part of the signature state which he sends to Charlie. By doing so,
Bob tries to increase his probability to later forge a message. In short,
if Bob alters the state in the multiport,
then Charlie's null-port of the multiport will occasionally (or more often) detect photons. By demanding that the protocol is aborted if too many photons are detected at the multiport null-port, Charlie limits the degree to which Bob can manipulate Charlie's
quantum signature state. The amount that this happens is bounded by using the relationship of fidelity with the trace distance as in~\cite{QDSPRL}. This reduces Bob's active attacks to (modified) passive forging attacks. The forging probability is then bounded by
\EQ{P(for)\leq \exp \left(-2\left(C'_{min}-s_v-\sqrt{\epsilon+r}\right)^2L\right)+2\exp (-2\epsilon^2L),\label{active forging}}
where $\epsilon$ is an arbitrarily positive constant, chosen when designing the protocol, and $C'_{min}$ is a lower bound for the minimum cost that Bob can achieve when making his best guess. $C'_{min}$ is calculated by allowing Bob to base his guess on a measurement  on coherent states with amplitude $\sqrt{3/2} \alpha.$ The minimum cost
for the optimal
measurement  $C_{min}'$ is smaller than $C_{min}$, since Bob now has access to a state with greater amplitude,
but can be bounded using the same method as we previously used to bound $C_{min}$.
The above expression holds, and it decays exponentially, provided that
\EQ{C'_{min}-s_v-\sqrt{\epsilon+r}>0.}

\subsection{Robustness}

After having analyzed the security of the protocol against repudiation and forging, we examine the robustness of the protocol. If all participants are honest, it should succeed with probability approaching one exponentially quickly as $L$ increases. Here we will first consider the probability of rejecting a message if all parties are honest and then the probability of aborting the protocol if all parties are honest. Both of them should, and do, decay exponentially.

If all parties are honest, then the probability that the phase $\theta_i$ is eliminated, when the state $\rho_i$ with phase $\theta_i$ is sent, is equal to the number of detections at the detector $D(\neg\theta_i)$ over the total number of pulses of $\rho_i$. This probability is the diagonal element $C_{i,i}$ of the cost matrix in eq. (\ref{exp cost matrix}). Therefore, averaged over all possible states, the probability that the correct state is ruled out is $p_h=1/4\sum_i C_{i,i}$. For Bob to fail to authenticate, we need that there are more than $s_a L$ mismatches. Using Hoeffding's inequalities, this occurs with probability
\EQ{P({hon~rej})\leq \exp \left(-2(s_a-p_h)^2L\right)}
provided that $s_a\geq p_h$, which can be made to decay exponentially in $L$.

On the other hand, abort can only occur if the number of the null-port counts exceed $r L$. It is interesting to note, that if we assume that Bob and Charlie are honest during the distribution stage, there is no need to keep count of the counts of the null-port and no need to abort the protocol. It is only due to active forging and the need to restrict the ability of Bob to tamper with the response state, that we require to consider aborting.
In order for our protocol to be correct, we need to choose $r$ to be larger, by at least some finite amount $\epsilon$, than the average number of null-port counts (dark-counts) when all parties are honest, which is $r_h=r-\epsilon$.  Then the probability for abort if all parties are honest decays exponentially with the signature length,
\EQ{P(hon~ab)  \leq \exp \left(-2\epsilon^2 L \right). \label{correctness}}

\subsection{Summary}

Here we summarize the performance of the protocol. In particular, if we
consider only passive forging, we have three main inequalities bounding the probabilities for undesired events, namely
\EQ{P({hon~rej})\leq \exp \left(-2(s_a-p_h)^2L\right)}
\EQ{P(for)\leq  \exp \left(-2 \left(C_{min}-s_v\right)^2 L\right),}
\EQ{P(rep)\leq \exp (-\frac{(s_v -s_a)^2}2 L)}
together with the conditions on the parameters,
\EQ{C_{min}\geq s_v\geq s_a\geq p_h.}
From this equation and the definition of the gap $g=C_{min}-p_h$, we can see that for positive gap, there exist choices of $s_a$ and $s_v$ that would make the protocol secure and robust. From Eq. (\ref{gap-adv}) it follows that the existence of a positive gap is guaranteed if the cost matrix of the experiment leads to a positive guaranteed advantage $guad>0$.
Note also, that as stated above, we do not need to keep track of the multiport null-port counts, leading to possible abort also if all participants are honest, unless we are also considering active forging. Nevertheless, according to \eqref{correctness}, the probability for abort if all participants are honest also decays exponentially in $L$.

We assume that we are equally interested in robustness,
security against repudiation and security against forging. By choosing $s_a=p_h+g/4$ and $s_v=p_h+3g/4$, the bounds for the forging, repudiation and honest rejection probability become equal. Since $g=C_{min}-p_h$, all three equations lead to
\EQ{P(for)=P(rep)=P(hon~rej)\leq \exp (-\frac{g^2}8 L).}
We see that with this choice of the parameters $s_a$ and $s_v$, we achieve exponential security against forging and repudiation, and exponentially small probability of abort if all parties are honest. The exponential decay rate depends on $g$, which for the demonstration provided in the main paper is found to be small.

There are two figures of merit by which a QDS scheme is evaluated. The first one is the length $L$ of the quantum signature that Alice needs to send to be able to securely sign
a ``half-bit'' message. The length $L$ is the QDS analogue of the ratio of raw key length to established secure key length in QKD. The second figure of merit is the rate by which signatures are distributed. This is the number of bits that Alice can sign per second, and is the clock rate divided by $L$.

For both of the above figures of merit, one needs to define what is meant by secure and robust, i.e. how small probability of repudiation, forging and honest rejection we desire. Statements about the performance of a QDS scheme should include the security level requested. In the estimates given below, we will require that the QDS scheme has a security level of $0.01\%$, which means that the probability of forging, repudiation or honest rejection are all less than $10^{-4}$.

Here we should also remark on how the length $L$ for a given security level depends on the losses in an experiment. Assume that we have a cost matrix coming from an experiment with given losses $l$, and that we want to extrapolate the cost matrix if the losses change to $l'$. Since the cost matrix elements are determined by the ratio of the number of detection events to the average number of pulses sent, having higher losses implies that the overall matrix is multiplied by $(1-l)/(1-l')$. If the guaranteed advantage of the experiment with losses $l$ is $guad$, then the guaranteed advantage with losses $l'$ will simply be $guad'=guad (1-l)/(1-l')$. From Eq. (\ref{gap-adv}) we can therefore see that the length $L$ required for a given security level increases quadratically as the transmitted signal (one minus losses) decreases.

In our experiment, we obtained a gap $g=1.20\times10^{-6}$ which for a security level of $0.01\%$ leads to a length $L=5.10\times10^{13}$ required to sign a ``half-bit''. This is clearly a very large number and to make a QDS scheme practical, one needs to improve it significantly. However, there are three changes that can be implemented that would indeed result in a significant improvement. We present them below along with our estimate of the expected performance of the corresponding experiments.

The first change is to use a different amplitude $\alpha$. The smaller the $\alpha$ is, the greater the value of $p_{min}$, which in its turn affects the value of the gap. On the other hand, the guaranteed advantage decreases with smaller $\alpha$. It is likely that there is an optimal value for $\alpha$, which is not in the range explored in our experiment. Using data from a subsequent experiment further exploring unambiguous state elimination,  for which we have a greater range of $\alpha$, and extrapolating using the known losses from the multiport, we have good indication that the optimal value of $\alpha^2$ is around $0.5$. Moreover, we estimate that this would lead to a significantly larger gap of $g=8.05\times10^{-5}$. This gives a length $L=1.14\times10^{10}$ for a security level of $0.01\%$.

The second potential improvement is to use a protocol that does not require the multiport, which is a major source of losses. Removing the multiport means that the security against repudiation needs to be guaranteed in an alternative way. We have recently proposed some QDS schemes that do not require the multiport \cite{No_multiport} and one could modify those protocols to use phase-encoded coherent states, similar to the current realization. We estimate that implementing all these change will result in a gap of $g=1.96\times10^{-4}$ and length $L=1.19\times10^{9}$ for a security level of $0.01\%$.

Finally, we could increase the clock rate, and therefore the transmission rate. The phase modulators, VCSEL, and driving electronics are capable of clock-rates up to 3.3~GHz. This would not change the length $L$, but it would reduce the required time to distribute the same number of secure signatures bits by a factor of 33, and therefore significantly improve the rate. In particular, we expect that making these changes will lead to distributing signatures at a rate capable of signing 1.4 bits per second with a security level of $0.01\%$.


\begin{thebibliography}{99}

\bibitem{QDS}
D. Gottesman and I.  Chuang,
{arXiv:quant-ph}/0105032v2 (2001).

\bibitem{ErikaOrig}
E. Andersson, M. Curty, and I. Jex,
{Phys. Rev. A} {\bf 74}, 022304  (2006).

\bibitem{OurNatComm}
P. J. Clarke {\it et al.},
{Nat. Commun.} {\bf 3}, 1174 (2012).

\bibitem{OurArXiv}
V. Dunjko, P. Wallden, and E. Andersson, Phys. Rev. Lett. {\bf 112} 040502 (2014).


\bibitem{stevebook}
S. Barnett, {\it Quantum Information}, Oxford University Press, pp 103-104 (2009).

\bibitem{OppenUSE}
S. Bandyopadhyay, R. Jain, J. Oppenheim, and C. Perry,
Phys. Rev. A {\bf 89}, 022336 (2014).

\bibitem{QKDreview}
N. Gisin, G. Ribordy, W. Tittel, and H. Zbinden,
{Rev. Mod. Phys.} {\bf 74}, 145--195 (2002).

\bibitem{QKDReviewSecurity}
V. Scarani {\em et al.},
{Rev. Mod. Phys.} {\bf 81}, 1301--1350 (2009).

\bibitem{RealQKD}
L. Widmer, {\it Cerberis: High-Speed Encryption with Quantum Cryptography} in {\it  Internet – Technical Development and Applications}, {Editors E. Tkacz and A Kapczynski}
{(Springer, Berlin, 2009)}, {Advances in Intelligent and Soft Computing} {\bf 64}, {217}.

\bibitem{TokyoQKD}
For example www.{\allowbreak}magi{\allowbreak}qtech{\allowbreak}.com, www.{\allowbreak}idquantique{\allowbreak}.com, www.{\allowbreak}quintessence{\allowbreak}labs{\allowbreak}.com, and www.{\allowbreak}sequrenet{\allowbreak}.com, all visited 12\textsuperscript{th} May 2014

\bibitem{qmem}
P. C. Maurer {\it et al.},
{Science} {\bf 336}, 6086 (2012).

\bibitem{qmemThewalt}
K. Saeedi {\it et al.},
{Science} {\bf 342}, 830 (2013).

\bibitem{Marand} 	
C. Marand and P. D. Townsend,
{Opt. Lett.} {\bf 20}, 1695--1697 (1995).

\bibitem{ExpUSD}
F. E. Becerra, J. Fan, and A. Migdall,
Nat. Commun. {\bf 4,} 2028 (2013).

\bibitem{Supp} See Supplementary Material for full details of the experimental methods, and the security statements and proofs.
................


\bibitem{GHzDetector}
P. J. Clarke {\em et al.},
{New J. Phys.} {\bf 13,} 075008 (2011).



\bibitem{Min-cost} P. Wallden, V. Dunjko and E. Andersson, J. Phys. A.: Math. Theor. {\bf 47}, 125303 (2014).


\bibitem{Arrazola}
J. M. Arrazola and N. L\"utkehnaus, arXiv:1309.5005 (2013).



\bibitem{RobustGHz}
P. J. Clarke, R. J. Collins, P. A. Hiskett, P. D. Townsend, and G. S. Buller,
{Appl. Phys. Lett.} {\bf 98,} 131103 (2011).


\bibitem{No_multiport} V. Dunjko, P. Wallden and E. Andersson, arXiv:1403.5551 (2014).

\bibitem{Jitter}
M. Wahl {\it et al.},
{Rev. Sci. Instrum.} {\bf 79}, 12113 (2008).



\bibitem{DaSilva}
T. F. Da Silva, G. B. Xavier, G. P. Tempor\~ao, and J. P. von der Weid,
{Opt. Express} {\bf 20}, 18911--24 (2012).



\bibitem{SPAD}
A. Spinelli, L. M. Davis, and H. Dautet,
{Rev. Sci. Instr.} {\bf 67,} 55 (1996).



\bibitem{Photodet1}
R. J. Collins, R. H. Hadfield, and G. S. Buller,
{J. Nanophot.} {\bf 4}, 040301 (2010).

\bibitem{Photodet2}
G. S. Buller and R. J. Collins,
{Meas. Sci. Technol.} {\bf 21}, 012002 (2010).




\bibitem{Panda}
A. Kumar and  A. Ghatak, {\it Polarization of light with applications in optical fibers},
SPIE Press, Bellingham, Washington (2011).






\bibitem{GPS}
L. Trigo and D. Slomovitz, {\it Rubidium atomic clock with drift compensation,}
in 2010 Conference on Precision Electromagnetic Measurements (CPEM) 472 -- 473 (IEEE, 2010).


\bibitem{QDSPRL} The Supplementary material of: V. Dunjko, P. Wallden, and E. Andersson,
Phys. Rev. Lett. {\bf 112}, 040502 (2014).



\bibitem{CavesPRA} C. Caves, C. Fuchs, and R. Schack
\textit{Phys. Rev. A} {\bf 66} 062111  (2002).

\bibitem{vanEnk} S. J. van Enk,
    \textit{Phys. Rev. A} {\bf 66} 042313 (2002)



\bibitem{Townsend1993} P. D. Townsend, J. G. Rarity, and P. R. Tapster, Electron. Lett. {\bf 29}, 1291 (1993).



\bibitem{hoeff}
 W. Hoeffding,
 {\em J. Amer. Statist. Assoc.},
  {\bf 58} 301, (1963).

\bibitem{Helstrom}
C. W. Helstrom,
\newblock {\em { Quantum detection and estimation theory}}.
\newblock Academic Press, New York, (1976).





\end{thebibliography}
 \end{document}